\documentclass[12pt]{iopart}

\usepackage{iopams} 
\usepackage{setstack}
\usepackage{mathrsfs}
\usepackage{epsfig}
\usepackage{graphicx}
\usepackage{bm}
\usepackage{cite}

\begin{document}

\title{Expansion of nanoplasmas and laser-driven nuclear fusion in single exploding clusters}

\author{F Peano$^{1}$, JL Martins$^{1}$, RA Fonseca$^{1,2}$, F Peinetti$^{3}$, R Mulas$^{3}$, G Coppa$^{3}$, I Last$^{4}$, J Jortner$^{4}$, and LO Silva$^{1}$}
\address{$^{1}$GoLP/Instituto de Plasmas e Fus\~ao Nuclear, Instituto Superior T\'ecnico, 1049-001 Lisboa, Portugal}
\address{$^{2}$Departamento de Ci\^encias e Tecnologias da Informa\c{c}\~ao, Instituto Superior de Ci\^encias do Trabalho e da Empresa, 1649-026 Lisboa, Portugal}
\address{$^{3}$Dipartimento di Energetica, Politecnico di Torino, 10129 Torino, Italy}
\address{$^4$School of Chemistry, Tel-Aviv University, Ramat Aviv, 69978 Tel-Aviv, Israel}
\eads{\mailto{fabio.peano@ist.utl.pt}, \mailto{luis.silva@ist.utl.pt}}

\date{\today}

\begin{abstract}
The expansion of laser-irradiated clusters or nanodroplets depends strongly on the amount of energy delivered to the electrons and can be controlled by using appropriately shaped laser pulses. In this paper, a self-consistent kinetic model is used to analyze the transition from quasineutral, hydrodinamic-like expansion regimes to the Coulomb explosion (CE) regime when increasing the ratio between the thermal energy of the electrons and the electrostatic energy stored in the cluster. It is shown that a suitable double-pump irradiation scheme can produce hybrid expansion regimes, wherein a slow hydrodynamic expansion is followed by a fast CE, leading to ion overtaking and producing multiple ion flows expanding with different velocities. This can be exploited to obtain intracluster fusion reactions in both homonuclear deuterium clusters and heteronuclear deuterium-tritium clusters, as also proved by three-dimensional molecular-dynamics simulations.
\end{abstract}


\submitto{\PPCF}
\maketitle

\newcommand{\ped}[1]{_{\mathrm{#1}}}
\newcommand{\api}[1]{^{\mathrm{#1}}}
\newcommand{\diff}[1]{\mathrm{d}#1}
\newcommand{\pped}[1]{_{\scriptscriptstyle #1}}
\newcommand{\aapi}[1]{^{\scriptscriptstyle #1}}
\newcommand{\dfrac}[2]{{\displaystyle\frac{#1}{#2}}}

\section{Introduction}

The interaction of ultraintense lasers with jets of molecular clusters or nanodroplets (with typical size in the range $1-100$ nm and containing $10^2-10^8$ particles) is a central research topic \cite{Krainov_rep}, with important applications, such as tabletop nuclear fusion for compact neutron sources \cite{Ditmire_Nature1,Ditmire_Nature2,fusion_exp1,fusion_exp2,fusion_exp3,fusion_exp4,fusion_exp5}, or the laboratory investigations of nucleosynthesis reactions, relevant to astrophysical scenarios \cite{Last_PRL_1,Heidenreich,Last_PRA_2}.

Clustered media can be regarded as sparse distributions of tiny solid targets, a peculiar configuration that allows for both a deep penetration of the laser radiation and a strong laser-matter coupling with many individual, overdense targets, thus providing extremely efficient energy absorption \cite{Ditmire_PRL_3}.
When hit by an ultraintense laser beam, the neutral atoms in a cluster are promptly ionized (cf. Ref. \cite{Last_JCP_1} for a detailed analysis of the concurring ionization mechanisms in different laser/cluster configurations) and a dense ``nanoplasma'' \cite{Ditmire_PRA_1, Krainov_rep} is formed. The free electrons then absorb energy from the laser pulse \cite{Mulser} and start expanding, causing the formation of strong electric fields, which lead to efficient ion acceleration, as first predicted by Dawson \cite{Dawson}. 
When the energy transferred to the electrons is much smaller than the electrostatic energy stored in the ion core, charge separation is localized to regions much smaller than the cluster \cite{Ditmire_PRA_1,Fukuda,Mora1,Mora2,Mora3,Murakami}, which then remains quasi-neutral and undergoes a hydrodynamic-like expansion \cite{hydro1,hydro2,hydro3,hydro4,hydro5}; in opposite conditions (e.g. with small deuterium clusters exposed to extremely intense laser radiation) the electrons suddenly escape from the cluster and the remaining bare-ion distribution undergoes a pure Coulomb explosion (CE) \cite{Kaplan_PRL}. 
In intermediate situations, the expansion dynamics is a mixture of the phenomenology of the two limits, with the expansion process being strongly dependent on the self-consistent dynamics of ions and trapped electrons \cite{Peano_PRL_2,Murakami}. When increasing the laser energy, or when lowering the cluster size and density, the expansion conditions vary smoothly from quasi-neutral, hydrodynamic-like regimes to pure CE regimes, as confirmed by particle-in-cell (PIC) simulations \cite{Kishimoto_POP,Peano_PhD} of the self-consistent laser-cluster interaction and by kinetic or fluid modeling of the expansion of finite-size, non-quasi-neutral plasma expansions \cite{Peano_PRL_2,Peano_PRE,Murakami}.
Therefore, the expansion regime can be controlled by regulating the amount of energy transferred to the electrons \cite{Peano_PoP}, which can be obtained with appropriately shaped laser beams.
An important example is the irradiation of homonuclear deuterium clusters with two sequential laser pulses having different intensities \cite{Peano_PRL_1}. In this way, one can taylor the expansion dynamics so as to induce overrunning between ions and the consequent formation of expanding shells containing multiple ion flows. Within such structures (here denoted as ``shock shells'', following the terminology in \cite{Kaplan_PRL}), the relative velocities between deuterium ions from different flows can be sufficiently high for energetic collisions and \textit{dd} fusion reactions to occurr \cite{Peano_PRA}. Since these intracluster reactions occur early in the expansion, and before different exploding clusters overlap, they are expected to produce a time resolved burst of fusion neutrons before the bulk neutron signal due to intercluster reactions \cite{Kaplan_PRL,Peano_PRA}. In the case of heteronuclear clusters, e.g. deuterium-tritium clusters, ion species having different charge-to-mass ratios expand with different velocities making intracluster, interspecies reactions possible also with standard single-pulse irradiation \cite{Last_PRA_2,Li}.

This paper provides an organic review of the work on controlled expansions of clusters and nanoplasmas published in \cite{Peano_PRL_2,Peano_PRE,Peano_PoP,Peano_PRL_1,Peano_PRA}, complemented with novel simulation results obtained with a recently developed molecular-dynamics technique \cite{Last_PRA}. In the following, the transition to the CE regime is analyzed with a self-consistent kinetic model (section \ref{sec:expansion}), the concept of shock shell is briefly reviewed (section \ref{sec:shocks}), and the possibility of achieving intracluster nuclear reactions in homonuclear clusters \cite{Kaplan_PRL,Peano_PRA} by driving a slow expansion followed by a sudden CE \cite{Peano_PRL_1,Peano_PRA} is explored (section \ref{sec:reactions}).

\section{Kinetic analysis of spherical plasma expansions}
\label{sec:expansion}

In order to analyze the influence of the electrons on the expansion dynamics, a kinetic model for the collisionless expansion of a spherically symmetric nanoplasma has been developed \cite{Peano_PRL_2}, based on the assumptions that the electrons are nonrelativistic and resorting to the large mass disparity between electrons and ions. 
The expansion process is divided in two stages: an initial charging transient with frozen ions, and the long term expansion of both ions and electrons.
In the second stage, the expansion dynamics is described self-consistently by following the radial motion of the cold ions, while accounting for the three-dimensional dynamics of the hot electrons using a sequence of ``ergodic'' equilibrium configurations \cite{Peano_PRE}, represented by stationary solutions of the Vlasov equation that depend on the total energy $\epsilon=m{\bf v}^2/2+\Phi(r)$ only. The electron density instantaneously in equilibrium with the electrostatic potential $\Phi$ can be written as $
n\ped{e}\left(r;\{\Phi\}\right) =\frac{1}{4\pi r^2}\int\rho\left(\epsilon\right)
\mathscr{P}\left(r,\epsilon;\{\Phi\}\right)\diff\epsilon$
where $\{\Phi\}$ indicates functional dependence on $\Phi$, $\rho$ is the energy distribution of the electrons, and
$\mathscr{P}\left(r,\epsilon;\{\Phi\}\right) = r^2\left( \epsilon+e\Phi\right)^{\frac{1}{2}} / \int {r^{\prime}}^2
\left[ \epsilon+e\Phi\left(r^{\prime}\right)\right]^{\frac{1}{2}}\diff r^{\prime}$
is the probability density of finding an electron having total energy $\epsilon$ at the radial position $r$. The self-consistent potential $\Phi$ satisfies the nonlinear Poisson equation $\nabla^2\Phi = 4\pi e\left[n\ped{e}\left(r;\{\Phi\}\right)-n\ped{i}\right]$. By resorting to the theory of adiabatic invariants for time varying Hamiltonians \cite{Peano_PRE,Ott,Grismayer}, 
a closed set of equations is then obtained in the form
\numparts
\begin{eqnarray}
	&M \dfrac{\partial^2 r\ped{i}}{\partial t^2} = -Ze \dfrac{\partial \Phi}{\partial r}(r\ped{i})
	\label{eq:model_a}\\
	&\dfrac{1}{r^2}\dfrac{\partial}{\partial r}\left(r^2\dfrac{\partial \Phi}{\partial r}\right) = 4\pi
	e\left(n\ped{e}-Zn\ped{i}\right)
	\label{eq:model_b}\\
	&n\ped{i}(r\ped{i}) = n\ped{i0}(r_0)\dfrac{r_0^2}{r\ped{i}^2}\!\Big/\dfrac{\partial r\ped{i}}{\partial r_0}
	\label{eq:model_c}\\
	&n\ped{e} = \dfrac{1}{4\pi r^2}\displaystyle\int \rho_0(\epsilon_0)
	\mathscr{P}\left(r,\epsilon;\{\Phi\}\right)\diff{\epsilon_0}
	\label{eq:model_d}\\
	&\dfrac{\diff\epsilon}{\diff t} = -e\displaystyle\int \dfrac{\partial\Phi}{\partial t} 	
	\mathscr{P}\left(r,\epsilon;\Phi\right)\diff r
	\label{eq:model_e}
\end{eqnarray}
\endnumparts
where $M$ and $Ze$ are the ion mass and charge, respectively. Equations \eref{eq:model_a} and \eref{eq:model_e}, determining the ion trajectory $r\ped{i}(r_0,t)$ ($r_0$: initial radius) and the evolution of the electron energy $\epsilon(\epsilon_0,t)$ ($\epsilon_0$: initial energy), are coupled with the nonlinear Poisson equation \eref{eq:model_b}. In Eq. \eref{eq:model_d}, the relation $\rho(\epsilon,t)\diff{\epsilon}=\rho_0(\epsilon_0)\diff{\epsilon_0}$ has been used, where $\rho_0(\epsilon_0)$ is the energy distribution of the electrons resulting from the initial charging transient, a fast process that cannot be described by Equations \eref{eq:model_a}--\eref{eq:model_e} \cite{Peano_PRE,Peano_PoP}. The self-consistent shape of $\rho_0(\epsilon_0)$ is determined by a virtual charging transient, in which the ions stay immobile, while an external potential barrier, initially confining the electrons, is gradually moved from $R_0$ to infinity, with a series of small radial displacements (the validity of this procedure has been confirmed by particle-in-cell simulations \cite{Peano_PRL_2} and ad-hoc solutions of the VP model \cite{Peano_PRE}). The initial, nonequilibrium electron distribution is assumed to be a Maxwellian, with temperature $T_0$, so that the whole expansion dynamics is fully determined by the single dimensionless parameter $\hat{T}_0 = Zk\ped{B}T_0/\epsilon\ped{CE}=3\lambda\ped{D0}^2/R_0^2$ ($\epsilon\ped{CE}=Ze^2N_0/R_0$: maximum ion energy attainable from the CE of a uniform ion sphere with radius $R_0$ and total charge $eN_0$; $\lambda\ped{D0}$: initial electron Debye length), which accounts for both the initial electron temperature and the cluster size and density.
\begin{figure}[!htbp]
\centering \epsfig{file=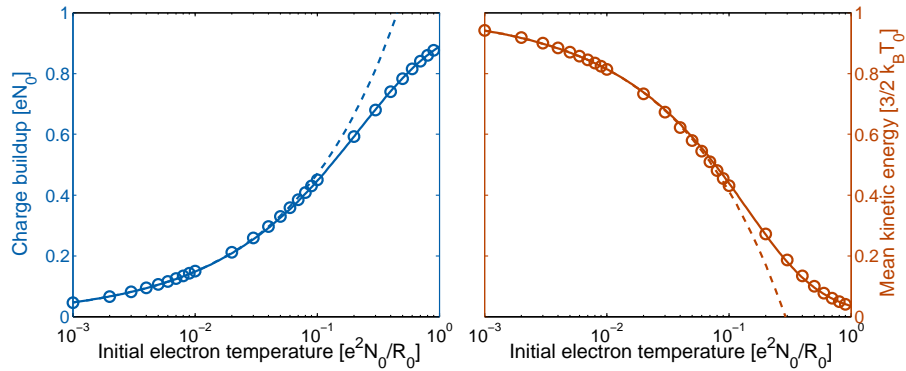, width=4.75in}
\caption{(Color online) Charge buildup $\Delta Q$ (left panel) and mean kinetic energy of trapped electrons, $\mathcal{E}$, (right panel) as functions of $\hat{T}_0$, after the charging transient with immobile ions. Circles refer to the ergodic model, solid lines to the fit laws in the text. Dashed lines show the power-law behaviors for $\hat{T}_0 \ll 1$.}
\label{fig:Qeq_Teq}
\end{figure}
The positive charge buildup at the ion front, $\Delta Q$, and the mean kinetic energy of the trapped electrons, $\mathcal{E}$, as functions of $\hat{T}_0$, are displayed in Fig. \ref{fig:Qeq_Teq}. Simple fits for $\Delta Q$ and $\mathcal{E}$ are found as $\Delta Q = eN_0 \mathcal{F}_{2.60}(\sqrt{6/\mathrm{e}}\hat{T}_0^{1/2})$ and 
$\mathcal{E} = \frac{3}{2}k\ped{B}T_0[1-\mathcal{F}_{3.35}(1.86\hat{T}_0^{1/2})]$, where
$\mathcal{F}_{\mu}(x) = x/(1+x^{\mu})^{1/\mu}$. For $\hat{T}_0 \ll 1$ (i.e. $\lambda\ped{D0} \ll R_0$), the fits reduce to $\Delta Q/(eN_0) \simeq (6\hat{T_0}/\mathrm{e})^{1/2}$, thus recovering the theoretical results for planar expansions \cite{Mora1,Crow}, and to $\mathcal{E}/(\frac{3}{2} \ k\ped{B}T_0) \simeq 1 - 1.86\:\hat{T}_0^{1/2}$.

\begin{figure}[!htbp]
\centering \epsfig{file=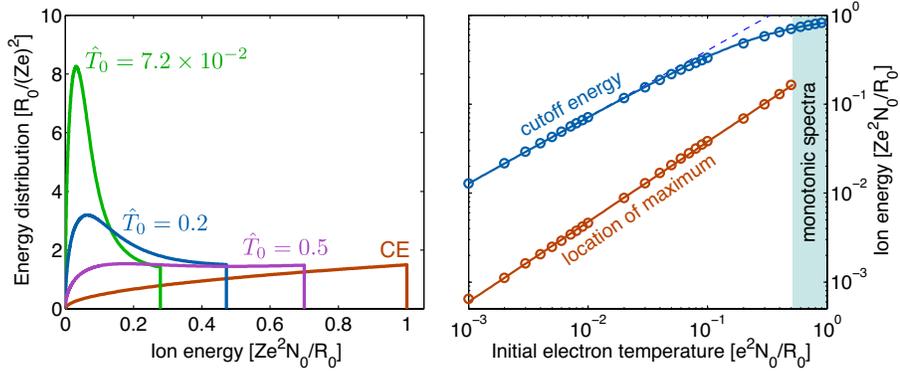, width=4.75in}
\caption{(Color online) Asymptotic ion energy spectra for different values of $\hat{T}_{0}$, compared with the theoretical asymptotic spectrum for the uniform CE case, $3R_0/(2Z^2e^2) \ (\epsilon/\epsilon\ped{CE})^{1/2}$ (left panel); cutoff ion energy and location of the maximum in the ion energy spectrum as functions of $\hat{T}_0$ (right panel). Circles refer to the ergodic model, solid lines to the fit laws in the text. The dashed line represents the power-law behavior of $\epsilon\ped{max}$ for $\hat{T}_0 \ll 1$.}
\label{fig:ionspectrum_T0}
\end{figure}
When the initial equilibrium is reached and the ions are allowed to expand, gaining kinetic energy, the electrons cool down and the charge buildup decreases (until a ballistic regime is reached for both species \cite{Manfredi}). Depending on the value of $\hat{T}_0$, such behavior strongly affects the ion dynamics and the resulting distribution of ion energy \cite{Peano_PRL_2,Peano_PRE}: for $\hat{T}_0 > 0.5$, the spectrum is monotonic as in a CE; for $\hat{T}_0 < 0.5$, it exhibits a local maximum far from the cutoff energy, thus being qualitatively different from the CE case. In this respect, the transition value $\hat{T}_0 = 0.5$ can be considered as the lower bound for the validity of the CE approximation. The cutoff energy is fit by $\epsilon\ped{max}=\mathcal{F}_{1.43}(2.28\:\hat{T}_0^{3/4})\epsilon\ped{CE}$, 
which, for $\hat{T_0}\ll 1$, reduces to $\epsilon\ped{max}\simeq 2.28\:\hat{T}_0^{3/4}\epsilon\ped{CE}$; for $\hat{T}_0 < 0.5$, the location of the maximum is fit by the power-law $\epsilon\ped{peak}=0.3\hat{T}_0^{0.9}\epsilon\ped{CE}$. These scaling laws, valid for any combination of $R_0$, $n_0$, and $T_0$, can be useful to interpret experimental data \cite{Peano_PRL_2,Sakabe1,Sakabe2,Sakabe3}. 

\section{Shock shells in nonuniform Coulomb explosions}
\label{sec:shocks}

In the absence of electrons, the explosion dynamics of a spherical distribution of cold ions is described by Equation \eref{eq:model_a}, which reduces to $M\partial^2r\ped{i}/\partial t^2 = Z^2e^2 N\ped{i}(r\ped{i},t)/r\ped{i}^2$,
where $N\ped{i}(r,t)$ is the number of ions enclosed by a sphere of radius $r$ at time $t$.

If the initial profile of ion density is uniform and step-like (i.e. $n\ped{i0}$ for $r \leqslant R_0$), the repulsive electric field grows linearly for $r<R_0$, reaching its maximum at the outer boundary, which causes the outer ions to be always faster than the inner ones. Thus, ions never overtake each other, $N\ped{i}(r,t)$ is conserved along the ion trajectories (so is the total energy of each ion), and the equation of motion can be integrated analytically \cite{Kaplan_PRL,Parks,Li_PRA,Kovalev_PoP}, yielding, for ions initially at rest, the expansion velocity $v\ped{i} = [2(r\ped{i}-r_0)/(3r\ped{i})]^{1/2}r_0\omega\ped{pi}$, 
where $\omega\ped{pi}=(4\pi Z^2e^2n\ped{i0}/M)^{1/2}$ is the initial ion plasma frequency. The radial trajectory is then given, in implicit form, by $[\xi\ped{i} \left(\xi\ped{i}-1\right)]^{1/2} + \log[\xi\ped{i}^{1/2}+(\xi\ped{i}-1)^{1/2}] = (2/3)^{1/2}\omega\ped{pi}t$, 
where $\xi\ped{i}=r\ped{i}/r_0$ is the expansion factor. In this solution, $\xi\ped{i}$ is independent of $r_0$, the $v-r$ phase-space profile is always a straight line with equation $v=[2(\xi-1)/(3\xi^3)]^{1/2}\omega\ped{pi}r$, and the ion density maintains its step-like form, decreasing in time as $n\ped{i0}\xi(t)^{-3}$. The asymptotic ion energy distribution is $3R_0/(2Z^2e^2) \ (\epsilon/\epsilon\ped{CE})^{1/2}$ (cf. Fig. \ref{fig:ionspectrum_T0}).

\begin{figure}[!htbp]
\centering \epsfig{file=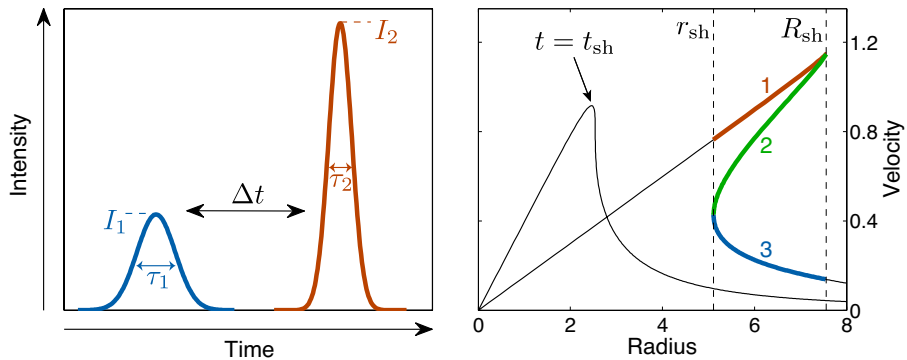, width=4.75in}
\caption{(Color online) Sketch of the double-pump irradiation scheme (left panel) and example of the typical formation and evolution of a shock shell with three-branched profile, the shock shell, in the $v-r$ phase space (right panel): early in the explosion, the phase-space profile bends on its right until it becomes multi-valued at time $t=t\ped{sh}$ and an expanding shock shell forms, with inner and outer boundaries $r\ped{sh}$ and $R\ped{sh}$, respectively. Energetic collisions can occur between ions belonging to different velocity branches.}
\label{fig:DP}
\end{figure}

If, more realistically, nonuniform density profiles are considered, the expansion features change qualitatively \cite{Peano_PRL_1,Peano_PRA}: when the initial density is a decreasing function of $r$, the repulsive Coulomb field reaches its maximum within the ion sphere \cite{Kaplan_PRL,Peano_PhD}, leading to ion overtaking and to the formation of multiple-flow regions (shock shells) with characteristic multi-branched phase-space profile as in Fig. \ref{fig:DP} (since the ion trajectories are no longer independent of one another, the above analytical solution no longer holds \cite{Kaplan_PRL,Peano_PhD,Kovalev_PoP}). The physical interest of these shock shells resides in the appearance of multiple flows with large relative velocities within a single exploding cluster, which can lead to energetic ion-ion collisions and to intracluster reactions in homonuclear clusters \cite{Kaplan_PRL, Peano_PRL_1,Peano_PRA}.
An effective strategy to produce large-scale shock shells in a controlled fashion is combining different expansion regimes so that a slow hydrodynamic-like expansion, providing a smoothly decreasing ion-density profile, is followed by an abrupt CE. This is achievable with a double-pump irradiation scheme as in Fig. \ref{fig:DP}, wherein a weak pulse (intensity: $I_1$; duration: $\tau_1$) is followed by a strong pulse (intensity: $I_2>I_1$; duration: $\tau_2$), with suitable time delay $\Delta t$: $I_1$ must be sufficiently high to ionize the atoms, creating a nanoplasma, but not so high as to expel a significant fraction of the electrons from the cluster, whereas $I_2$ must be high enough to drive a sudden CE. In these conditions, a pronounced shock shell is formed, whose features are determined by the key double-pump parameters, namely $I_1$ and $\Delta t$, which strongly affects the density profile of the cluster when being hit by the second laser. The effectiveness of the technique has been demonstrated \cite{Peano_PRL_1,Peano_PRA,Peano_PoP} by resorting to two- and three-dimensional PIC simulations performed using the OSIRIS 2.0 framework \cite{OSIRIS}, closely matching realistic physical scenarios.

\section{Intracluster nuclear reactions}
\label{sec:reactions}
The ability of producing multiple flows with high relative velocities within single exploding clusters makes the phenomenon attractive as a possible way to induce intracluster fusion reactions in homonuclear deuterium clusters. As pointed out in \cite{Kaplan_PRL}, the rates for intracluster reactions can be significantly higher than those for intercluster reactions, because the typical densities within a shock shell ($\lesssim n\ped{i0} \sim 10^{22}$ cm$^{-3}$) are higher than those within the hot plasma filament resulting from the exploded clusters ($\sim 10^{19}$ cm$^{-3}$). However, it is crucial to consider also that the expanding shock-shell stays appreciably dense only for a very brief time ($\sim 10$ fs), much shorter than the typical disassembly time of the plasma filament ($\sim 10-100$ ps): as shown in \cite{Peano_PRA}, the rates of intracluster reactions exhibit a sharp, time-resolved peak right after the shock-shell formation. 
Once the expansion dynamics is known, the number of intracluster fusion reactions is obtained by summing over all possible contributions from collisions between ions belonging to different velocity branches in a shock shell (cf. Fig. \ref{fig:DP}). The number of reactions per unit time and volume, $\mathcal{R}$, is given by
\begin{equation}
\mathcal{R} = \sum_{h<k}
                         n_{h}(r)
                         n_{k}(r)
                         \sigma \left( | v_{h} - v_{k}|  \right)
                         |v_{h} - v_{k}|
                         \mathrm{,}
\label{eq:rrate1}
\end{equation}
where $\sigma$ is the cross section for \textit{dd} fusion, while $n_{h}$ and $v_{h}$ indicate, respectively, the ion density and velocity on the $h$th branch. The number of intracluster fusion reactions, $\mathcal{N}$, is obtained by integrating $\mathcal{R}$ over time and space, as 
\begin{equation}
\mathcal{N}= 4\pi \int_{t\ped{sh}}^{+\infty} \!\!\!\! \int_{r\ped{sh}}^{R\ped{sh}} \!\! \mathcal{R} {r}^{2}\diff{r} \diff{t} \mathrm{,}
\label{eq:Nreact}
\end{equation} 
where $r\ped{sh}$, $R\ped{sh}$, and $t\ped{sh}$ are the shock-shell boundaries and formation time, respectively. In \cite{Peano_PRA}, the influence of the double-pump parameters on $\mathcal{N}$ was analyzed using a simple 1D model wherein the laser field of the first pulse gradually strips the initially neutral cluster of a part of its electrons according to a cluster barrier suppression ionization (CBSI) model (cf. Ref. \cite{Last_JCP_1}).                                                                                                                                                                                                                                                                                                                                                                                                                                                                                                                                                                                                                                                                                                                                                                                                                                                                                                                                                                                                                                                                  
This provided preliminary estimates for the optimal combinations of delay and intensities that maximize $\mathcal{N}$, and suggested that, for very large clusters ($R_0 \gtrsim 100$ nm), the intracluster reaction yield can become comparable with the intercluster neutron yield, with $\sim 10\%$ of the fusion reactions arising from intracluster collisions \cite{Peano_PRA}.

Accurate calculations of the intracluster reaction yields achievable with the double-pump irradiation of homonuclear deuterium clusters are currently being performed resorting to three-dimensional molecular-dynamics simulations based on the recently developed scaled electron and ion dynamics (SEID) technique \cite{Last_PRA,Last_new}, for different values of laser intensities, durations, and delays. 
Preliminary results indicate that, with $I_1$ in the range $10^{15}-10^{17}$ Wcm$^{-2}$, $\tau_1$ in the range $20-30$ fs, $I_2=10^{20}$ Wcm$^{-2}$, $\tau_2 = 20$ fs, and $\Delta t$ in the range $110-180$ fs, the number of intracluster reaction per cluster can be as high as $\mathcal{N}=4\times10^{-3}$ with a cluster size $R_0 = 110$ nm.
These values are comparable with the estimate in \cite{Peano_PRA}; furthermore, as predicted in \cite{Peano_PRA}, it is found that the optimal intensities of the first pulse are significantly lower, typically by an order of magnitude, than those estimated with the CBSI model. 
Calculations based on the SEID technique are being performed also in the case of heteronuclear deuterium-tritium clusters and nanodroplets irradiated by a single laser pulse, for either homogenous mixtures or layered targets composed by a deuterium core surrounded by a tritium shell. Owing to the higher value of the deuterium-tritium fusion cross section for the energy range considered, higher number of intracluster reactions, on the order of $\mathcal{N}=0.1-1$, are obtained with respect to the pure-deuterium case.
Detailed results of these numerical simulations will be presented in a future publication \cite{Last_new}.

\section{Summary}
The ergodic model has been used to simulate the collisionless expansion of spherical nanoplasmas driven by energetic electrons over a wide range of values of plasma size, plasma density, and initial electron thermal energy, combined in the single dimensionless parameter $\hat{T}_0$: a qualitative change in the shape of the asymptotic energy spectrum of the ions when approaching the Coulomb explosion regime has been identified and accurate fit laws for the relevant expansion features have been provided.
The douple-pump irradiation technique to induce the formation of multiple ion flows during the expansion of large homonuclear deuterium clusters or nanodroplets has been described, and its applicability to produce intracluster \textit{dd} and \textit{dt} fusion reactions has been proved resorting to three-dimensional numerical simulations performed with a recently developed scaled molecular dynamics technique.

\ack
Work partially supported by FCT (Portugal) through grants PDCT/POCI/66823/2006, SFRH/BD/39523/2007, and SFRH/BPD/34887/2007, and by the European Community - New and Emerging Science and Technology Activity under the FP6 "Structuring the European Research Area" programme (project EuroLEAP, contract number 028514). Part of the simulations discussed here were performed using the IST cluster (IST/Lisbon).

\end{document}